\def\~{{$\tilde{\phantom{a}}$}}

\documentclass [12pt] {article}
\usepackage{epsfig}
\usepackage{color}

\def\thebibliography#1{\section{References}\markboth
 {REFERENCES}{REFERENCES}\list
 {[\arabic{enumi}]}{\settowidth\labelwidth{[#1]}\leftmargin\labelwidth
 \advance\leftmargin\labelsep
 \usecounter{enumi}}
 \def\newblock{\hskip .11em plus .33em minus -.07em}
 \sloppy
 \sfcode`\.=1000\relax}
\def\upcite#1{\raise6pt\hbox{\scriptsize
\cite{#1}}}
  \def\lsim{\mathrel {\vcenter {\baselineskip 0pt \kern 0pt
    \hbox{$<$} \kern 0pt \hbox{$\sim$} }}}
    \def\gsim{\mathrel {\vcenter {\baselineskip 0pt \kern 0pt
    \hbox{$>$} \kern 0pt \hbox{$\sim$} }}}

\def\hline{\noalign{\hrule \vskip2pt}}

\def\|{\ifmmode\Vert\else \char`\|\fi}
\ifx\oldzeta\undefined                          
  \let\oldzeta=\zeta                            
  \def\zzeta{{\raise 2pt\hbox{$\oldzeta$}}}     
  \let\zeta=\zzeta                              
\fi

\ifx\oldchi\undefined                           
  \let\oldchi=\chi                              
  \def\cchi{{\raise 2pt\hbox{$\oldchi$}}}       
  \let\chi=\cchi                                
\fi

\def\frac#1#2{{#1 \over #2}}

\def\half{\ifinner {\scriptstyle {1 \over 2}}
   \else {1 \over 2} \fi}

\def\abs#1{\left\vert#1\right\vert}	

\def\simge{\mathrel{%
   \rlap{\raise 0.511ex \hbox{$>$}}{\lower 0.511ex \hbox{$\sim$}}}}
\def\simle{\mathrel{
   \rlap{\raise 0.511ex \hbox{$<$}}{\lower 0.511ex \hbox{$\sim$}}}}

\def\buildchar#1#2#3{{\null\!                   
   \mathop#1\limits^{#2}_{#3}                   
   \!\null}}                                    
\def\overcirc#1{\buildchar{#1}{\circ}{}}

\def\slashchar#1{\setbox0=\hbox{$#1$}           
   \dimen0=\wd0                                 
   \setbox1=\hbox{/} \dimen1=\wd1               
   \ifdim\dimen0>\dimen1                        
      \rlap{\hbox to \dimen0{\hfil/\hfil}}      
      #1                                        
   \else                                        
      \rlap{\hbox to \dimen1{\hfil$#1$\hfil}}   
      /                                         
   \fi}                                         %

\def\subrightarrow#1{
  \setbox0=\hbox{
    $\displaystyle\mathop{}
    \limits_{#1}$}
  \dimen0=\wd0
  \advance \dimen0 by .5em
  \mathrel{
    \mathop{\hbox to \dimen0{\rightarrowfill}}
       \limits_{#1}}}                           

\def\overlay#1#2{\ifmmode%
\setbox0=\hbox{$#1$}%
\setbox1=\hbox to\wd0{\hss$#2$\hss}\else%
\setbox0=\hbox{#1}%
\setbox1=\hbox to\wd0{\hss#2\hss}\fi%
#1\hskip-\wd0\box1 }

\def\pmb#1{\leavevmode\setbox0=\hbox{#1}%
\kern-.02em\copy0\kern-\wd0
\kern.04em\copy0\kern-\wd0
\kern-.02em\raise.04em\box0 }

\def\vereq#1#2{\lower3pt\vbox{\baselineskip1.5pt \lineskip1.5pt
\ialign{$\m@th#1\hfill##\hfil$\crcr#2\crcr\sim\crcr}}}

\def\tensor#1{\protect\@ontopof{#1}{\leftrightarrow}{1.15}\mathord{\box2}}
\def\overstar#1{\protect\@ontopof{#1}{\ast}{1.15}\mathord{\box2}}
\def\overdots#1{\protect\@ontopof{#1}{\cdots}{1.0}\mathord{\box2}}
\def\overcirc#1{\protect\@ontopof{#1}{\circ}{1.2}\mathord{\box2}}
\def\loarrow#1{\protect\@ontopof{#1}{\leftarrow}{1.15}\mathord{\box2}}
\def\roarrow#1{\protect\@ontopof{#1}{\rightarrow}{1.15}\mathord{\box2}}

\def\@ontopof#1#2#3{%
{\mathchoice
{\@@ontopof{#1}{#2}{#3}\displaystyle\scriptstyle}%
{\@@ontopof{#1}{#2}{#3}\textstyle\scriptstyle}%
{\@@ontopof{#1}{#2}{#3}\scriptstyle\scriptscriptstyle}%
{\@@ontopof{#1}{#2}{#3}\scriptscriptstyle\scriptscriptstyle}%
}%
}

\def\@@ontopof#1#2#3#4#5{%
\setbox0=\hbox{$#4#1$}%
\setbox1=\hbox{$#5#2$}%
\setbox2=\hbox{}\ht2=\ht0 \dp2=\dp0 %
\ifdim\wd0>\wd1 %
\setbox1=\hbox to\wd0{\hss\box1\hss}%
\mathord{\rlap{\raise#3\ht0\box1}\box0}%
\else   %
\setbox1=\hbox to.9\wd1{\hss\box1\hss}%
\setbox0=\hbox to\wd1{\hss$#4\relax#1$\hss}%
\mathord{\rlap{\copy0}\raise#3\ht0\box1}%
\fi
}%

\def\lambdabar{\protect\@lambdabar}
\def\@lambdabar{%
\relax
\bgroup
\def\@tempa{\hbox{\raise.73\ht0
\hbox to0pt{\kern.25\wd0\vrule width.5\wd0
height.1pt depth.1pt\hss}\box0}}%
\mathchoice{\setbox0\hbox{$\displaystyle\lambda$}\@tempa}%
{\setbox0\hbox{$\textstyle\lambda$}\@tempa}%
{\setbox0\hbox{$\scriptstyle\lambda$}\@tempa}%
{\setbox0\hbox{$\scriptscriptstyle\lambda$}\@tempa}%
\egroup
}

\def\corresponds{{\lower.2ex\hbox{=}}{\rm\kern-.75em^\triangle}}
\def\succsim{\succ\kern-.9em_\sim\kern.3em}
\def\precsim{\prec\kern-1em_\sim\kern.3em}
\def\slantfrac#1#2{\kern1em^{#1}\kern-.3em/\kern-.1em_{#2}}

\begin{document}

\begin{center}
{\Large\bf Polarization Dependence of Emissivity}
\\

\medskip

David J.~Strozzi
\\ 
{\sl Department of Physics, Massachusetts Institute of Technology, 
Cambridge, MA 02139}
\\
Kirk T.~McDonald
\\
{\sl Joseph Henry Laboratories, Princeton University, Princeton, NJ 08544}
\\
(April 3, 2000)
\end{center}

\section{Problem}

Deduce the emissive power of radiation of frequency $\nu$ into
vacuum at
angle $\theta$ to the normal to the surface of a good conductor
at temperature $T$, 
for polarization both parallel and perpendicular to the plane of emission.

\section{Solution}

The solution is adapted from ref.~\cite{Wolf} (see also \cite{Landau}), 
and finds application in
the calibration of the polarization dependence of detectors for cosmic
microwave background radiation \cite{Wollack,Herzog}.

Recall Kirchhoff's law of heat radiation (as clarified by Planck
\cite{Planck}) that
\begin{equation}
{P_\nu \over A_\nu} = K(\nu,T) = {h \nu^3 / c^2 \over e^{h \nu / k T} - 1},
\label{e1}
\end{equation}
where $P_\nu$ is the emissive power per unit area per unit frequency interval
 (emissivity) and 
\begin{equation}
A_\nu = 1 - {\cal R} = 1 - \abs{E_{0r} \over E_{0i}}^2
\label{e1a}
\end{equation} 
 is
the absorption coefficient $(0 \leq A_\nu \leq 1)$, $c$ is the speed of light,
$h$ is Plank's constant and $k$ is Boltzmann's constant.
Also recall the Fresnel equations of reflection that
\begin{equation}
\left. {E_{0r} \over E_{0i}} \right|_\perp = 
{\sin(\theta_t - \theta_i) \over \sin(\theta_t + \theta_i)},
\qquad
\left. {E_{0r} \over E_{0i}} \right|_\parallel = 
{\tan(\theta_t - \theta_i) \over \tan(\theta_t + \theta_i)},
\label{e2}
\end{equation}
where $i$, $r$, and $t$ label the incident, reflected, and transmitted waves,
respectively.

The solution is based on the fact that eq.~(\ref{e1}) holds separately
for each polarization of the emitted radiation, and is also independent of 
the angle of the radiation.  This result is implicit in
Planck's derivation \cite{Planck} of Kirchhoff's law of radiation, and is
stated explicitly in \cite{Reif}.  That law 
describes the thermodynamic equilibrium of radiation emitted and absorbed 
throughout a volume.  
The emissivity $P_v$ and the absorption coefficient $A_\nu$ can depend on
the polarization of the radiation and on the angle of the radiation, but the
definitions of polarization
parallel and perpendicular to a plane of emission, and of angle relative
to the normal to a surface element, are local, while the energy 
conservation relation $P_\nu = A_\nu K(\nu,T)$ is global.  
A ``ray" of radiation whose polarization can be described as parallel to the 
plane of emission is, in general, a mixture of
parallel and perpendicular polarization from the point of view of the 
absorption process.  Similarly, the angles of emission and absorption of a ray
are different in general.   Thus, the concepts of parallel and perpendicular 
polarization and of the angle of the radiation are not well
defined after integrating over the entire volume.
Thermodynamic equilibrium can exist only if a single spectral intensity function
$K(\nu,T)$ holds independent of polarization and of angle.

All that remains is to evaluate the reflection coefficients ${\cal R}_\perp$ and
${\cal R}_\parallel$ for the two polarizations at a vacuum-metal interface.
These are well known \cite{Wolf,Landau,Stratton}, but we derive them for
completeness.

To use the Fresnel equations (\ref{e2}), we need expressions for 
$\sin\theta_t$ and
$\cos\theta_t$.  The boundary condition that the phase of the wave be 
continuous 
across the vacuum-metal interface leads, as is well known, to the general 
form of Snell's law:
\begin{equation}
k_i \sin\theta_i = k_t \sin\theta_t,
\label{e3a}
\end{equation}
where $k = 2 \pi / \lambda$ is the wave number.  Then,
\begin{equation}
 \cos\theta_t = \sqrt{1 - {k_i^2 \over k_t^2} \sin^2\theta_i}.
\label{e3b}
\end{equation}

To determine the relation between wave numbers $k_i$ and $k_t$ in vacuum and 
in the conductor,
we consider a plane wave of angular frequency $\omega = 2 \pi \nu$ and
complex wave vector {\bf k},
\begin{equation}
{\bf E} = {\bf E}_0 e^{i({\bf k_t} \cdot {\bf r} - \omega t)},
\label{e4}
\end{equation}
which propagates in a conducting medium with dielectric constant $\epsilon$,
permeability $\mu$, and conductivity $\sigma$.  The wave equation for the
electric field in such a medium is (in Gaussian units)
\begin{equation}
\nabla^2 {\bf E} - {\epsilon \mu \over c^2} {\partial^2{\bf E} \over
\partial t^2} = {4 \pi \mu \sigma \over c^2} {\partial{\bf E} \over
\partial t},
\label{e5}
\end{equation}
where $c$ is the speed of light.  We find the dispersion relation for the
wave vector $k_t$ on inserting eq.~(\ref{e4}) in eq.~(\ref{e5}):
\begin{equation}
k_t^2 = \epsilon \mu {\omega^2 \over c^2}  + i {4 \pi \sigma \mu \omega \over
c^2}.
\label{e6}
\end{equation}
For a good conductor, the second term of eq.~(\ref{e6}) is much larger than
the first, so we write
\begin{equation}
k_t \approx {\sqrt{2 \pi \sigma \mu \omega} \over c} (1 + i)
= {1 + i \over d} = {2 \over d(1 - i)},
\label{e7}
\end{equation}
where
\begin{equation}
d = {c \over \sqrt{2 \pi \sigma \mu \omega}} \ll \lambda
\label{e8}
\end{equation}
is the frequency-dependent skin depth.  Of course, on setting $\epsilon = 1 = \mu$ and 
$\sigma = 0$ we
obtain expressions that hold in vacuum, where $k_i = \omega/c$.

We see that for a good conductor $\abs{k_t} \gg k_i$, so according to eq.~(\ref{e3b})
we may take  $\cos\theta_t \approx 1$
to first order of accuracy in the small ratio $d/\lambda$.  Then the first of 
the Fresnel equations becomes
\begin{equation}
\left. {E_{0r} \over E_{0i}} \right|_\perp = 
{\cos\theta_i \sin\theta_t / \sin\theta_i - 1 \over 
 \cos\theta_i \sin\theta_t / \sin\theta_i + 1}
= {(k_i / k_t) \cos\theta_i   - 1 \over 
 (k_i / k_t) \cos\theta_i + 1}
\approx {(\pi d / \lambda)(1 - i) \cos\theta_i - 1 \over 
 (\pi d / \lambda)(1 - i) \cos\theta_i + 1},
\label{e9}
\end{equation}
and the reflection coefficient is approximated by
\begin{equation}
{\cal R}_\perp = \abs{E_{0r} \over E_{0i}}^2_\perp  
\approx 1 - {4 \pi d \over \lambda} \cos\theta_i
= 1 - 2 \cos\theta_i\sqrt{\nu \over \sigma}.
\label{e10}
\end{equation}
For the other polarization, we see that
\begin{equation}
\left. {E_{0r} \over E_{0i}} \right|_\parallel = 
\left. {E_{0r} \over E_{0i}} \right|_\perp
{\cos(\theta_i + \theta_t) \over \cos(\theta_i - \theta_t)}
\approx \left. {E_{0r} \over E_{0i}} \right|_\perp
{\cos\theta_i - (\pi d / \lambda)(1 - i) \sin^2\theta_i \over 
 \cos\theta_i + (\pi d / \lambda)(1 - i) \sin^2\theta_i},
\label{e11}
\end{equation}
so that
\begin{equation}
{\cal R}_\parallel \approx {\cal R}_\perp  
\left(1 - {4 \pi d \over \lambda} {\sin^2\theta_i \over \cos\theta_i} \right)
\approx 1 - {4 \pi d \over \lambda \cos\theta_i}
= 1 - {2 \over \cos\theta_i} \sqrt{\nu \over \sigma} .
\label{e13}
\end{equation}
An expression for 
${\cal R}_\parallel$ valid to second order in $d/\lambda$ has been given in
ref.~\cite{Stratton}.  For $\theta_i$ near $90^\circ$, ${\cal R}_\perp \approx
1$, but eq.~(\ref{e13}) for ${\cal R}_\parallel$ is not accurate.  Writing
$\theta_i = \pi/2 - \vartheta_i$ with $\vartheta_i \ll 1$, eq.~(\ref{e11}) 
becomes
\begin{equation}
\left. {E_{0r} \over E_{0i}} \right|_\parallel \approx 
{\vartheta_i  - (\pi d / \lambda)(1 - i) \over 
 \vartheta_i + (\pi d / \lambda)(1 - i) },
\label{e11a}
\end{equation}
For $\theta_i = \pi/2$, ${\cal R}_\parallel = 1$, and 
${\cal R}_{\parallel,\rm min} = (5 - \sqrt{2}) / (5 + \sqrt{2}) = 0.58$ for
$\vartheta_i = 2 \sqrt{2} \pi d / \lambda$.

Finally, combining eqs.~(\ref{e1}), (\ref{e1a}), (\ref{e10}) and (\ref{e13}) 
we have
\begin{equation}
P_{\nu\perp} \approx {4 \pi d \cos\theta \over \lambda^3} 
{h \nu \over e^{h \nu / k T} - 1},
\qquad
P_{\nu\parallel} \approx 
{4 \pi d \over \lambda^3 \cos\theta} 
{h \nu \over e^{h \nu / k T} - 1},
\label{e14}
\end{equation}
and
\begin{equation}
{P_{\nu\perp} \over P_{\nu\parallel}} = \cos^2\theta
\label{e15}
\end{equation}
for the emissivities at angle $\theta$ such that $\cos\theta \gg d/\lambda$.  

The conductivity $\sigma$ that appears in eq.~(\ref{e14}) can be taken as the
dc conductivity so long as the wavelength exceeds 10 $\mu$m \cite{Wolf}.  If
in addition $h \nu \ll k T$, then eq.~(\ref{e14}) can be written
\begin{equation}
P_{\nu\perp} \approx {4 \pi d\ k T \cos\theta \over \lambda^3}, \qquad
P_{\nu\parallel} \approx  {4 \pi d\ k T \over \lambda^3 \cos\theta },
\label{e16}
\end{equation}
in terms of the skin depth $d$.

We would like to thank Matt Hedman, Chris Herzog and Suzanne Staggs for
conversations about this problem.


\begin{thebibliography}{99}

\bibitem{Wolf}
M.~Born and E.~Wolf,
{\em Principles of Optics}, 7th ed.,
(Cambridge U.\  Press, Cambridge, 1999), sec.~14.2.

\bibitem{Landau}
L.D.~Landau and E.M.~Lifshitz,
{\em The Electrodynamics of Continuous Media}
(Pergamon Press, Oxford, 1960), sec.~67.

\bibitem{Wollack}
E.J.~Wollack,
{\em A measurement of the degree scale cosmic background radiation anisotropy
at 27.5, 30.5, and 33.5 GHz},
Ph.D.\ dissertation (Princeton University, 1994), Appendix C.1.1.

\bibitem{Herzog}
C.~Herzog,
{\sl Calibration of a Microwave Telescope},
Princeton U.\ Generals Expt.\ (Oct.\ 26, 1999).

\bibitem{Planck}
M.~Planck,
{\em The Theory of Heat Radiation}
(Dover Publications, New York, 1991), chap.~II, especially sec.~28.

\bibitem{Reif}
F.~Reif,
{\em Fundamentals of statistical and thermal physics}
(McGraw-Hill, New York, 1965), sec.~9.14.

\bibitem{Stratton}
J.A.~Stratton,
{\em Electromagnetic Theory}
(McGraw-Hill, New York, 1941), sec.~9.9.

\end{thebibliography}
\end{document}